\documentclass[conference]{IEEEtran}
\IEEEoverridecommandlockouts

\usepackage{threeparttable}
\usepackage{cite}
\usepackage{svg}
\usepackage{amsmath,amssymb,amsfonts}
\usepackage{listings}
\usepackage{xcolor}
\usepackage{tcolorbox}
\usepackage{algorithmic}
\usepackage{graphicx}
\usepackage{subfigure}
\usepackage{textcomp}
\usepackage{xcolor}
\usepackage[a4paper, total={184mm,239mm}]{geometry}
\usepackage{booktabs}
\usepackage{multirow}
\usepackage{makecell}
\usepackage{ulem}
\usepackage{xurl}
\usepackage[linesnumbered,ruled,vlined]{algorithm2e}

\newcommand{\github}{\url{https://github.com/AutoBench/CorrectBench}}
\def\BibTeX{{\rm B\kern-.05em{\sc i\kern-.025em b}\kern-.08em
    T\kern-.1667em\lower.7ex\hbox{E}\kern-.125emX}}
\usepackage{stfloats} 

\newcommand{\projectname}{{CorrectBench}}
\newcommand{\figname}{{Fig.}}
\newcommand{\tabname}{Table}

\newcommand{\py}{Python}
\newcommand{\vl}{Verilog}

\definecolor{vgreen}{RGB}{104,180,104}
\definecolor{vblue}{RGB}{49,49,255}
\definecolor{vorange}{RGB}{255,143,102}

\SetCommentSty{mycommfont}
\SetKwInput{KwInput}{Input}                
\SetKwInput{KwOutput}{Output}              
\SetKwInput{KwData}{Modules}              

\begin{document}

\title{\projectname: Automatic Testbench Generation with Functional Self-Correction using LLMs for HDL Design\vspace{-0.5cm}
}


\author{
\IEEEauthorblockN{Ruidi Qiu$^1$,
Grace Li Zhang$^2$,
Rolf Drechsler$^3$,
Ulf Schlichtmann$^1$,
Bing Li$^4$}
\IEEEauthorblockA{$^1$Technical University of Munich, $^2$TU Darmstadt, $^3$University of Bremen, 
$^4$University of Siegen
}
\IEEEauthorblockA{Email: \{r.qiu, ulf.schlichtmann\}@tum.de, grace.zhang@tu-darmstadt.de, drechsler@uni-bremen.de}
bing.li@uni-siegen.de\vspace{-0.6cm}
}

\maketitle

\begin{abstract}

Functional simulation is an essential step in digital hardware design. Recently, there has been a growing interest in leveraging Large Language Models (LLMs) for hardware testbench generation tasks. However, the inherent instability associated with LLMs often leads to functional errors in the generated testbenches. Previous methods do not incorporate automatic functional correction mechanisms without human intervention and still suffer from low success rates, especially for sequential tasks. 
To address this issue, we propose \projectname{}, an automatic testbench generation framework with functional self-validation and self-correction. Utilizing only the RTL specification in natural language, the proposed approach can validate the correctness of the generated testbenches with a success rate of 88.85\%. Furthermore, the proposed LLM-based corrector employs bug information obtained during the self-validation process to perform functional self-correction on the generated testbenches.
The comparative analysis demonstrates that our method achieves a pass ratio of 70.13\% across all evaluated tasks, compared with the previous LLM-based testbench generation framework's 52.18\% and a direct LLM-based generation method's 33.33\%.
Specifically in sequential circuits, our work's performance is 62.18\% higher than previous work in sequential tasks and almost 5 times the pass ratio of the direct method.
The codes and experimental results are open-sourced at the link: \github .

\end{abstract}

\renewcommand\IEEEkeywordsname{Keywords}
\begin{IEEEkeywords}
    Large Language Models, HDL Design, Hardware Simulation, Testbench Generation.
\end{IEEEkeywords}
\vspace{-0.3cm}

\section{Introduction}

Simulation-based functional verification, relying on a testbench (TB), is among the most prevalent verification techniques employed during the initial phases of hardware design. The engineering effort required to design a testbench for functional simulation remains significantly high \cite{coverage_testgen_review}, with much of this effort being task-specific. This specificity complicates finding a generic method to optimize the process.
Previous works, such as those by \cite{coverage_testgen2, mcellin2022avert, kitchen2007stimulus}, have primarily focused on automating the generation of test stimuli for the design under test (DUT), which constitutes the front end of the functional simulation. The back end involves verifying the correctness of the signals from the DUT, which is highly specialized, making traditional automation methods ineffective and thus unattainable for the fully automated testbench design.

The increasing application of LLMs in the digital hardware design process suggests an alternative approach to automating testbench design. Recent studies \cite{chipchat, LLMforHLSxu, romewasnot_hierachical, chipgpt, verilogeval,9927393,sun2024classificatio} demonstrated the effectiveness of LLMs in various aspects of hardware design, particularly in Register-transfer level (RTL) design. 
Some research efforts have extended beyond basic RTL design correction using LLMs  \cite{tsai2023rtlfixer, autochip}. In the realm of functional simulation-based verification, preliminary efforts have been made.
For instance, \cite{llmtbgenfsm} investigates the potential of LLMs in generating testbenches for finite state machines (FSMs), while \cite{autobench} introduces a framework called AutoBench, the first systematic and generic testbench generation framework. 
Although achieving an average 57\% improvement compared with directly generating testbench using LLMs, AutoBench still suffers from a low success rate. This limitation arises from the inherent uncertainty of LLMs, such as hallucination \cite{hallucinationllm} and laziness \cite{lazy_llm}. Additionally, AutoBench employs only syntax self-checking, similar to RTLFixer \cite{tsai2023rtlfixer}, without implementing functional self-checking. This is a common issue in current LLM-based hardware design methodologies, the absence of a self-checking mechanism indeed limits the potential performance of the AutoBench framework.
\begin{figure}
    \centering
    \includegraphics[scale=0.88]{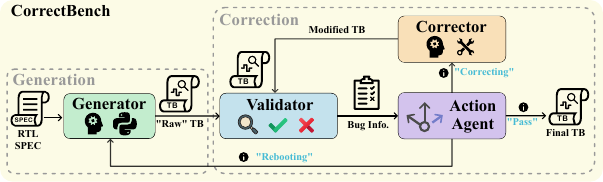}
    \vspace{-0.6cm}
    \caption{The outline of \projectname{} workflow.}
    \vspace{-0.6cm}
    \label{fig: CorrectBench workflow}
\end{figure}

To address the aforementioned issues, this paper proposes \projectname{}, the first framework for automatic testbench generation that incorporates functional self-validation and self-correction. Our framework utilizes the design specification (SPEC) of the device under test (DUT) in natural language as the sole input, as illustrated in \figname{} \ref{fig: CorrectBench workflow}, while expanding the boundaries of current testbench generation methods. 
The contributions of this work are summarized as follows:

\begin{itemize}
    \item An \textbf{action-based} testbench self-validation and self-correction \textbf{framework} is proposed. The total testbench generation pass ratio is improved up to 70.13\%, compared with 52.18\% in the previous work and 33.33\% in a direct method where LLMs are applied directly to generate test benches. Specifically in sequential circuits, our work's performance is 62.18\% higher than previous work in sequential tasks and almost 5 times the direct method.
    \item A scenario-based testbench \textbf{self-validator} is proposed, validating the correctness of the generated testbench via a particular matrix. The validator only takes the task specification in natural language as the input information and achieves an average \textbf{88.85\%} validation accuracy.
    \item An LLM-based testbench \textbf{self-corrector} is used to take the bug information from the validator as the input. The corrector makes a \textbf{34.33\%} contribution in the total improvement compared with previous work.
    \item The code, dataset, and experimental results are open-sourced on \github.
\end{itemize}

\section{Background and Motivation}

\subsection{AutoBench: Automatic Testbench Generation Framework}
\label{subsec: 2-1 autobench}
\begin{figure}
    \centering
    \includegraphics[scale=0.93]{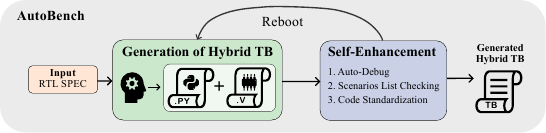}
    \vspace{-0.3cm}
    \caption{The outline of AutoBench workflow \cite{autobench}. AutoBench is used as the testbench generator in \figname{} \ref{fig: CorrectBench workflow}.
    }
    \vspace{-0.2cm}
    \label{fig: AutoBench workflow}
\end{figure}
\begin{figure}
    \centering
    \includegraphics[scale=0.73]{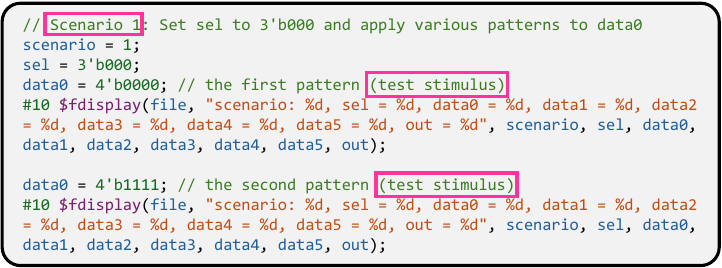}
    \vspace{-0.6cm}
    \caption{A demo of the test scenario and test stimuli in AutoBench's Verilog driver. In this demo, two stimuli are contained in one scenario. The output signals from DUT will be exported and checked by a Python checker later.}
    \vspace{-0.4cm}
    \label{fig: scenario/stimuli demo autobench}
\end{figure}
AutoBench \cite{autobench} is the first systematic and generic LLM-based testbench generation workflow. 
This workflow consists primarily of three components: the Verilog driver track, the Python checker track, and the simple self-enhancement stages, as illustrated in Figure \ref{fig: AutoBench workflow}. The framework's sole input is the RTL specification in natural language. Initially, the driver track generates a list of test scenarios and subsequently produces the Verilog driver, which drives the DUT to generate output signals under these scenarios. A test scenario is characterized by a specific set of test stimuli, as shown in \figname{} \ref{fig: scenario/stimuli demo autobench}. Subsequently, the checker track produces a Python checker. The Python checker is a Python code that generates the reference signals of the testbench and checks the correctness of DUT's output signals. The integration of the driver and checker constitute the hybrid testbench, which is further refined through self-enhancement stages, including syntax debugging, code completion, and scenario completion.

A significant challenge AutoBench faces is that it cannot check the correctness of the generated testbenches. 
The inherent instability of LLMs often leads AutoBench to fail in tasks that it is capable of solving. Although AutoBench includes a syntax debugging stage to correct the syntax of generated testbenches, it still lacks a mechanism 
to calibrate the generated testbenches, thus leading to a low pass rate.

\subsection{Motivation}
\label{subsec: 2-2 iterating llm design process}

To address the challenges above, self-validation and self-correction mechanisms can be incorporated into LLM-based testbench generation.
Similar but simpler strategies have been applied to the previous LLM-based hardware design.
For instance, RTLFixer \cite{tsai2023rtlfixer} implements syntactic checking and correction as preliminary efforts in this direction. However, it is only effective for syntax errors and thus has the same limitations as AutoBench.
Another research direction is AutoChip \cite{autochip}, which employs human-written testbenches to simulate the generated RTLs and uses testbench reports to inform the subsequent generation process.
While such feedback workflows prove effective, they typically rely on supplementary human-crafted content, such as testbenches, which contradicts the goal of a full automation process. 
Another study \cite{llmtbgenfsm} tries to use the DUT to evaluate the testbench's coverage and refine the testbench according to the coverage report. However, it can only partially assess coverage since the DUT's correctness is not inherently ensured.  

To enhance the quality of testbenches generated by LLMs, we propose \projectname{} with a functional self-validation and self-correction mechanism that surpasses basic syntactic checking and correction. 
In \projectname{}, a group of LLM-generated imperfect RTLs is used as the judge for the validation stage. These generated imperfect RTLs will be simulated, the results of which will be exploited to validate the correctness of the testbench generated by LLMs. Consequently, our self-validation module can provide a high validation success rate while needing no additional human-crafted content, providing higher flexibility in the practical testbench design process. Moreover, the conversation-based corrector will make full use of the bug information from the validator to perform an effective self-correction.

\section{Methodology}
\label{sec: methodology}

\subsection{Framework of \projectname}
\label{subsec 3-1: Framework and Action Agent of CorrectBench}

The framework of \projectname{} is drawn in \figname{} \ref{fig: CorrectBench workflow}, and is described in Algorithm \ref{alg: action agent} (in the next page). This work mainly focuses on the functional validation and correction of LLM-generated testbenches. Thus, the AutoBench \cite{autobench}, as shown in \figname{} \ref{fig: AutoBench workflow}, is used as the testbench generator of the proposed framework. 
The generated testbench, called ``raw'' TB, is sent to the validator to do the functional validation (blue box in \figname{} \ref{fig: CorrectBench workflow}). After validation, a report with correct, wrong, and uncertain test scenario indexes (bug information), as well as the correctness of the testbench, is provided to the action agent (purple box). The action agent then decides one of the three actions as the next action: correcting such testbench with the corrector (orange box), rebooting the whole process, or simply ending it. 

\normalem


  
    


\begin{algorithm}[!t]
\small
\DontPrintSemicolon
  
    \KwInput{DUT's Specification: \textit{SPEC}}
    \KwOutput{Final TestBench: $\textbf{TB}_{\textbf{final}}$}
    \KwData{Generator $\textbf{F}_\textbf{g}$, Validator $\textbf{F}_\textbf{v}$, Corrector $\textbf{F}_\textbf{c}$}
    $I_C \leftarrow 0$, $I_R \leftarrow 0$ \tcp*{initialize counters}
    $\textbf{A} \leftarrow ``\textit{None}"$  \tcp*{initialize the Action Agent}
    $\textbf{TB} \leftarrow \textbf{F}_{\textbf{g}}(\textit{SPEC})$ \tcp*{generate TB at the beginning}
    \While{$\textbf{A} \neq ``\textit{Pass}"$ }
    {
        $C_{\textit{TB}}, {\textit{Bugs}} \leftarrow \textbf{F}_\textbf{v}(\textbf{TB})$ \tcp*{validate TB, record TB correctness and bug information}
        \If{$(C_{\textit{TB}}=False)$ \textit{and} $(I_C < I^{max}_{C})$}
        {   
            $\textbf{A} \leftarrow ``\textit{Correcting}"$ \tcp*{Action: Correcting} 
            $I_C \leftarrow I_C + 1$ \\
            $\textbf{TB} \leftarrow \textbf{F}_{\textbf{c}}(\textbf{TB}, \textit{Bugs})$
        }
        \ElseIf{$(C_{\textit{TB}}=False)$ \textit{and} $(I_R < I^{max}_{R})$}
        {
            $\textbf{A} \leftarrow ``\textit{Rebooting}"$ \tcp*{Action: Rebooting}
            $I_R \leftarrow I_R + 1$ \\
            $I_C \leftarrow 0$ \tcp*{reset $I_C$ for a new rebooting iteration}
            $\textbf{TB} \leftarrow \textbf{F}_{\textbf{g}}(\textit{SPEC})$       
        }
        \Else 
        {
            \tcp{No error detected, or exceed max iteration} 
            $\textbf{A} \leftarrow ``\textit{Pass}"$ \tcp*{Action: Pass}
        }
    }
    $\textbf{TB}_{\textbf{final}} \leftarrow \textbf{TB}$

\caption{The workflow of \projectname{} }

\label{alg: action agent}

\end{algorithm}
\ULforem 

As shown in Algorithm \ref{alg: action agent} line 6, if the validator determines the testbench is wrong, the agent will first try to correct it with bug information by calling the corrector. If the correction iteration exceeds $I^{max}_C$, the following action will become ``rebooting'', which will go back to the generator and reset other parameters, such as the correction iteration, as is depicted in line 10. If the rebooting time exceeds the max value $I^{max}_R$, the whole system will give up, and the following action will be ``pass'', as shown in line 15. 
In the experiments, $I^{max}_C$ was set to 3 and $I^{max}_R$ was set to 10.

\subsection{Design of Scenario-Based Validator}
\label{subsec 3-2: validator}
The design of the validator in our work aims to 
accurately determine whether the testbench generated by the LLM is correct or not, given only the RTL specification without any additional information. If the testbench contains functional errors, the validator needs to provide as much information as possible to assist the subsequent corrector to locate and then correct the errors.

\subsubsection{Validation Methodologies}
\label{subsubsec: RS matrix}
The testbench generated by AutoBench includes multiple test scenarios as shown in \figname{} \ref{fig: scenario/stimuli demo autobench}., which are used in conjunction with the Python checker to evaluate whether the DUT can generate correct or erroneous outputs. 
Due to the instability in the LLM, the testbench's Python checker may generate erroneous reference signals in specific test scenarios (the definition of Python checker is mentioned in Section \ref{subsec: 2-1 autobench}). These wrong scenarios mean the testbench contains errors.
To validate whether there are actually such scenarios, an intuitive idea is to simulate a correct RTL design to compare its golden outputs and those described in the generated testbenches. However, this method is not viable since, at this stage, we only have the design specification. Although the LLM can also generate the RTL design with the design specifications, the correctness of this RTL design cannot be guaranteed.

To address the challenge described above, we use the LLM to generate a group of ``imperfect'' RTL designs, which might contain errors.
Since LLMs generate these RTL designs according to the correct design specifications, their errors tend to be randomly distributed due to the uncertainty of the LLM. Accordingly, it is unlikely for most RTL designs to have the same mistakes in the exact scenarios.  
Based on this analysis, we will simulate the RTL designs generated by the LLM with the testbench generated by AutoBench and collect the output correctness/errors of each scenario in the testbench. Assume that the number of generated RTLs and test scenarios are $N_{R}$ and $N_{S}$, respectively. An $N_{R} \times N_{S}$ boolean matrix can thus be obtained where 0/1 in the $i$th row and $j$th column represents the output of $j$th scenario in the testbench is wrong/correct according to the simulation result of the $i$th RTL design. We call this matrix \textit{RTL-Scenario matrix} (\textit{RS matrix}). In this work, $N_{R}$ is set to 20, and $N_{S}$ is set by the generator according to the task complexity.


\begin{figure}
    \centering
    \includegraphics[width=0.47\textwidth]{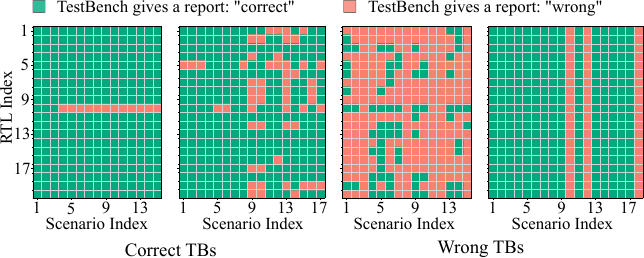}
    \vspace{-0.3cm}
    \caption{Examples of RS Matrices. The red/green color in the $i$th row and $j$th column represents the output of the $j$th scenario in the testbench is wrong/correct according to the simulation result of the $i$th RTL design. The two matrices on the left represent the correct TBs, whereas the matrix on the right indicates errors.}
    \vspace{-0.5cm}
\label{fig: RTL-Scenario Matrix}
\end{figure}

Examples of such RS matrices are illustrated in 
\figname{} \ref{fig: RTL-Scenario Matrix}. These matrices are generated from the experiments in Section \ref{subsec 4-3: Comparison of validators}. In this figure, a matrix row denotes the testbench's correctness report for an RTL with respect to all the test scenarios. The color red indicates that the testbench has a ``wrong'' output in a test scenario when an RTL design is used. 
On the contrary, a green block means ``correct'' outputs in the scenario when an RTL design is used. Similarly, a matrix column denotes the testbench for all RTLs in one test scenario.

In generating a group of RTL designs with the LLM, if an RTL design contains syntax errors, any associated reports on the output correctness in all the test scenarios will be discarded. If more than half of the RTL designs contain syntax errors, the system will regenerate the corresponding number of RTL designs until at least half of them are free from syntax errors. This approach ensures that sufficient information is available to validate the testbench accurately.
\subsubsection{Validation Criterion}
\label{subsubsec: validator criterion}
Though derived from imperfect RTLs, the RS matrix already provides information to determine the correctness of the generated testbench.
A simple criterion is to check the correctness of each column, which corresponds to a test scenario with all the generated RTL designs. 
If a column is completely red, indicating all the RTL designs generate an output different from that described in the testbench for a test scenario, 
there is a high possibility that the testbench itself contains mistakes for this scenario. 
Accordingly, a naive validation criterion can be used: 
if there is a column in the RS matrix that is completely red, we assume the corresponding scenario is wrong, and the testbench is wrong. 
This criterion is called \textbf{100\%-wrong}.

However, the above criterion is conservative when identifying erroneous scenarios in testbenches, resulting in a significant number of testbenches that are actually incorrect but mistakenly validated as correct.
Therefore, a stricter criterion is proposed. If 70\% of the RTL designs generate simulation results that are different from that described in the testbench for one test scenario, such test scenarios are marked as wrong, and the testbench is marked as wrong. 
The new criterion inevitably increases the risk of incorrectly classifying correct testbenches as erroneous. To alleviate this problem, an additional rule is applied based on the new criterion: if more than 25\% of the RTL designs completely match the testbench, indicating these RTLs are checked as correct across all scenarios (represented as an entirely green row in the RS matrix), then the testbench will be directly considered correct.
With the new rule about rows, the criterion of 70\% is called \textbf{70\%-wrong}. This criterion is finally chosen as the \projectname 's validation criterion

\subsection{Design of Corrector}
\label{subsec 3-3: Corrector}

\begin{figure}
    \centering
    \includegraphics[scale=0.54]{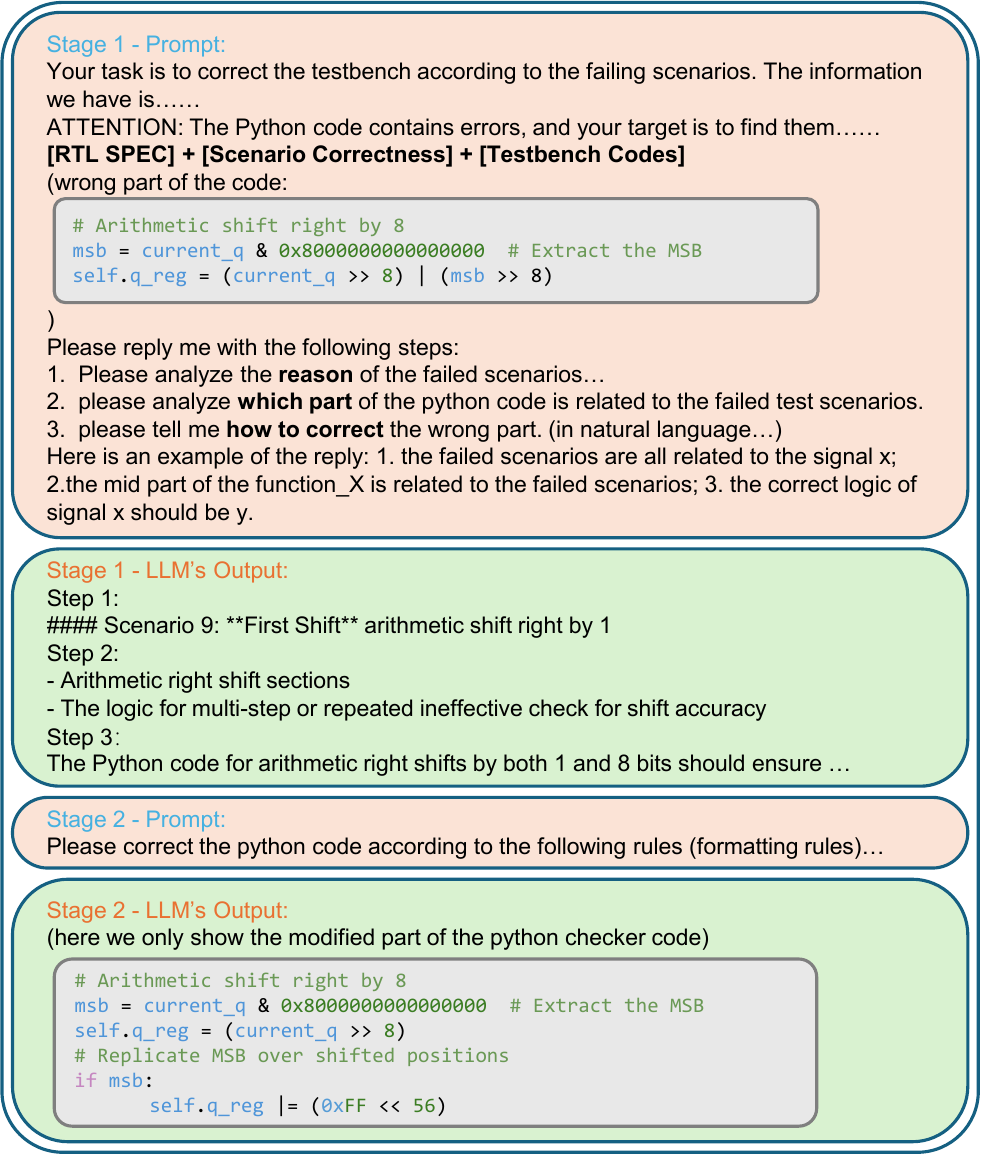}
    \vspace{-0.4cm}
    \caption{A Demo of Corrector. The RTL problem is \textit{shift18}, an arithmetic shifter. Some details are omitted to save space.}
    \label{fig: corrector demo}
    \vspace{-0.5cm}
\end{figure}
The corrector is a conversational stage based on an LLM, utilizing the model's reasoning capabilities.
When the validator detects errors in the testbench for at least one scenario, and if the correction iteration has not exceeded the maximum, the information obtained during the validation process is passed to the corrector for correction. The corrector can access the following information: the \textit{design specification} of the RTL, the testbench \textit{code}, the definition of each \textit{scenario} in testbench, and the \textit{indexes} of scenarios that are \textit{wrong}, \textit{correct}, or \textit{uncertain}. This scenario information from the previous step is crucial for error correction, as it helps the corrector more accurately pinpoint the location of the errors in the testbench code.

A heuristic chain of thought is employed to guide the LLM step by step in attributing existing error content with the aforementioned information. The whole correction is divided into two stages.

\subsubsection{Stage 1 - Reasoning}
LLM is guided to answer three questions -- \textit{why}, \textit{where}, and \textit{how}. A simplified demo is shown in \figname{} \ref{fig: corrector demo}. 
The first question directs the LLM to attribute the underlying causes of errors,
aiming to identify the root causes of errors, as there may be one fundamental cause for multiple wrong scenarios.
Building upon the analysis of the first question, the second question further directs the LLM to identify the location of the error in the testbench code.
Finally, the LLM will be directed to propose natural language-based methods for resolving these errors based on the source and location of testbench mistakes.
\subsubsection{Stage 2 - Correction}
With the information derived above, the LLM will be guided in modifying the testbench code. Additionally, the testbench code format is provided at this stage to prevent misformatting. Only the core code needs to be generated; the other codes, such as the fixed code interface, will be completed by a Python script. A demo of stage 2 is also shown in \figname{} \ref{fig: corrector demo}.

\section{Experimental Results}
\label{sec: experimental results}

\begin{table*}[t]
    \caption{Main results of proposed \projectname{} and comparison with other work.}
    \vspace{-0.4cm}
    \begin{center}
    
    \tabcolsep=0.4cm
    \begin{tabular}{cc*{2}{|ccc}}
        \toprule
        \multirow{3}*{Group} & \multirow{3}*{Metric} & \multicolumn{3}{c}{Ratio (\%)} & \multicolumn{3}{c}{\#Tasks}
        \\ 
        \cmidrule(lr){3-5} \cmidrule(lr){6-8}   
        & & \projectname{} & AutoBench \cite{autobench} & Baseline & \projectname{} & AutoBench & Baseline
        \\
        \midrule
        \multirow{3}*{\thead{Total\\(156)}}& Eval2 & 70.13\% (+36.80\%)*   & 52.18\% (+18.85\%)  & 33.33\%     & 109.4 (+57.4)     & 81.4 (+29.4)   & 52.0  \\
          & Eval1 & 79.49\% (+39.49\%)   & 57.05\% (+17.05\%)  & 40.00\%     & 124.0 (+61.6)    & 89.0 (+26.6)   & 62.4  \\
          & Eval0 & 99.87\% (+34.87\%)   & 94.62\% (+29.62\%)  & 65.00\%    & 155.8 (+54.4)   & 147.6 (+46.2)  & 101.4  \\
        \midrule
        \multirow{3}*{\thead{CMB\\(81)}}& Eval2 & 84.20\% (+30.62\%)   & 69.14\% (+15.56\%)  & 53.58\%     & 68.2 (+24.8)   & 56.0 (+12.6)   & 43.4  \\
          & Eval1 & 86.67\% (+27.66\%)   & 69.38\% (+10.37\%)  & 59.01\%     & 70.2 (+22.4)    & 56.2 (+8.4)   & 47.8  \\
          & Eval0 & 99.75\% (+19.50\%)   & 90.86\% (+10.61\%)  & 80.25\%    & 80.8 (+15.8)    & 73.6 (+8.6)   & 65.0  \\
        \midrule
        \multirow{3}*{\thead{SEQ\\(75)}}& Eval2 & 54.93\% (+43.46\%)   & 33.87\% (+22.40\%)  & 11.47\%     & 41.2 (+32.6)    & 25.4 (+16.8)   & 8.6  \\
          & Eval1 & 71.73\% (+52.26\%)   & 43.73\% (+24.26\%)  & 19.47\%     & 53.8 (+39.2)    & 32.8 (+18.2)   & 14.6  \\
          & Eval0 & 100.0\% (+51.47\%)   & 98.67\% (+50.14\%)  & 48.53\%    & 75.0 (+38.6)    & 74.0 (+37.6)   & 36.4  \\
        \bottomrule
    \end{tabular}
    \begin{tablenotes}
        \item[1] \hspace{0.3cm} * The values in parentheses represent the improvement of the method compared with the baseline.
    \end{tablenotes}
    \label{table: main results}
    \vspace{-0.8cm}
    \end{center}
\end{table*}
\subsection{Experimental Setup}
\label{subsec 4-1: Experimental Setup}
\subsubsection{Software Environment}
In this work, \textit{Icarus Verilog} \cite{IVerilog} was chosen as the \vl{} simulator. This is the most popular open-source \vl{} simulator, which also supports \textit{IEEE1800-2012} standards, including System Verilog syntax. All the \py{} codes were executed on \textit{Python 3.12.4 64-bit}. All the scripts and hardware simulations are run on servers with 2.40 GHz Xeon Silver 4314 or 2.60 Xeon Gold 6126 processors. The operating system is Linux.

\subsubsection{LLM Selection}
All the experiments in Section \ref{subsec 4-2: Main Results} and \ref{subsec 4-3: Comparison of validators}  were conducted on OpenAI's latest flagship model \textit{gpt-4o-2024-08-06}.
To demonstrate the compatibility of \projectname, we extended our evaluation in Section \ref{subsec 4-4: Performance on Other LLMs} to include Anthropic's flagship model, \textit{claude-3-5-sonnet-20240620}, and OpenAI's latest lightweight model, \textit{gpt-4o-mini-2024-07-18}. 

\subsubsection{Dataset}
This work uses the same dataset as AutoBench \cite{autobench}, extended from VerilogEval-Human \cite{verilogeval}. The extension includes mutant codes from the golden RTLs, which will only be used to evaluate the performance of \projectname. The dataset consists of 156 Verilog problems from HDLBits \cite{HDLBits}, including 81 combinational (CMB) problems and 75 sequential (SEQ) problems.

\subsubsection{Evaluation Criteria}
\label{evaluation criteria}

\begin{table}[t]\scriptsize
    \caption{Definitions of Evaluation Criteria in AutoEval \cite{autobench}}
    \centering
    \begin{tabular}{c|l}
        \toprule
        \textbf{\footnotesize{Type}} & \textbf{\footnotesize{Definition}} \\
        \midrule
        \midrule
        Failed & codes have syntax error \\
        \cmidrule{1-2}
        Eval0 & codes have no syntax error \\
        \cmidrule{1-2}
        Eval1 & codes passed Eval0; report \textit{passed} with the golden RTL code as DUT \\
        \cmidrule{1-2}
        \multirow{2}*{Eval2} & codes passed Eval1; use mutants of golden RTL as DUTs; have the\\ & same report as the golden testbench (\textit{passed} or \textit{failed}) \\
        \bottomrule
    \end{tabular}
    \label{tab: Evals}
    \vspace{-0.3cm}
\end{table}
In this study, AutoEval \cite{autobench} is utilized to conduct an evaluation of our proposed work. AutoEval includes three testbench evaluation criteria from syntactic to exhaustive, as is shown in \tabname{} \ref{tab: Evals}. The last criterion \textit{Eval2} utilizes 10 mutant RTLs as Design Under Test (DUTs) and compares the testbench's report (\textit{Failed} or \textit{Passed}) with the golden testbench. If its reports are the same as the golden testbench's on 80\% of the mutants, then the testbench will be recognized as ``Eval2 passed".

\subsection{Main Results}
\label{subsec 4-2: Main Results}

\subsubsection{Main Results}
To evaluate the performance of the proposed methodology, comparative experiments were conducted to show the performance of the proposed work against the previous work ``AutoBench" \cite{autobench} and the baseline of directly asking LLM to generate the testbench. In each experiment, we applied the testbench generation method to 156 tasks. To account for variability, we repeated each experiment five times.

The results of the comparison experiments are shown in \tabname{} \ref{table: main results}. The first column \textit{Group} shows the group of tasks sorted by circuit type. The second column \textit{Metric} denotes the evaluation criterion, as is discussed in Section \ref{evaluation criteria}. Columns 3 to 5 represent the performance of the testbench generation methods in the testbench pass rate, while columns 6 to 8 are the average number of passed ones among 156 tasks. 

As discussed in Section \ref{evaluation criteria}, the metric Eval 2 is the final evaluation criterion and is utilized as the testbench pass ratio to the testbench generation methods. For the total 156 tasks, columns 3, 4, and 5 in row 3 of \tabname{} \ref{table: main results} indicate that our \projectname{} outperforms both the baseline method and the previous AutoBench framework. Compared with AutoBench, our \projectname{} generates 34.40\% ($\frac{70.13\%}{52.18\%}-1$) more correct testbenches. In addition, our \projectname{} achieves more than two times ($\frac{70.13\%}{33.33\%}$) testbench Eval2 pass ratio on average than the Baseline's. This huge improvement is mainly from the sequential circuit tasks. 

In the previous work, the sequential tasks were quite challenging due to the higher complexity compared with combinational circuits, thus lowering the total pass ratio of the methods. Although AutoBench generates almost three times the correct testbenches than the baseline (col 4 and 5 in row 9, 33.87\% compared to 11.47\%), it still does not have a good performance in terms of the absolute numbers. Thanks to the collaboration of self-validator and self-corrector, our work achieves a pass ratio of 54.93\% for sequential circuits, which is 66.18\% higher (col 3 and 4 in row 9, $\frac{54.93\%}{33.87\%}-1$) than AutoBench and almost 5 times (col 3 and 5 in row 9, $\frac{54.93\%}{11.47\%}$) of the baseline method. This improvement marks a significant stride towards the practical applicability of our work.

\subsubsection{Contributions of Validator and Corrector}
\label{EXP main results contribution of validator and Corrector}
\begin{table}[t]
    \caption{Contributions of Validator and Corrector.}
    \vspace{-0.4cm}
    \begin{center}
    
    \resizebox{0.9\linewidth}{!}{
    \begin{tabular}{c|cc|c|cc}
        \toprule
        Group & \projectname{} & AutoBench & Gain & Val. & Corr.
        \\
        \midrule
        Total & 109.4 & 81.4 & 28.0 & 26.8 & 9.2
        \\
        CMB & 68.2 & 56.0 & 12.2 & 12.6 & 3.6
        \\
        SEQ & 41.2 & 25.4 & 15.8 & 14.2 & 5.6
        \\
        \bottomrule
    \end{tabular}
    }
    \label{table: contribution val corr}
    \end{center}
    \vspace{-0.6cm}
\end{table}
Compared to prior research, \projectname{} demonstrates substantial improvement by introducing automatic validation and correction. We conducted a comprehensive analysis to assess the contributions of the two primary strategies of our work, the validator and the corrector. This evaluation involved quantifying the average number of Eval2-passed tasks by using each strategy, as is shown in \tabname{} \ref{table: contribution val corr}. The item \textit{``Gain''} denotes the improvement of \projectname{} against the previous work AutoBench. The items \textit{``Val.''} and \textit{``Corr.''} denote the \projectname 's average Eval2 pass number where the validator or the corrector plays a significant role. Note that the preliminary step in calling the corrector is to call the validator first. Thus, the number 26.8 in column 5 already includes 9.2 in column 6, and the same cases are for groups CMB and SEQ.

Obviously, the number of \projectname 's Gain 28.0 (109.4 - 81.4) is almost equivalent to 26.8, the task number passed with validators, considering the results fluctuation of separately running \projectname{} and AutoBench. 
This means the enhancements observed in our \projectname{} can be primarily attributed to the newly involved functional checking mechanism. Among the 26.8 tasks successfully passed using validators, 34.33\% ($\frac{9.2}{26.8}$) tasks were achieved by applying the corrector, indicating that the corrector plays a significant role in our study. The SEQ group derives greater benefits from the corrector than the CMB group due to the increased complexity of SEQ, necessitating more thorough correction rather than simply applying ``rebooting'' action.

\subsection{Comparison of Different Validation Criteria}
\label{subsec 4-3: Comparison of validators}
\begin{figure}
    \centering
    \includegraphics[scale=0.28]{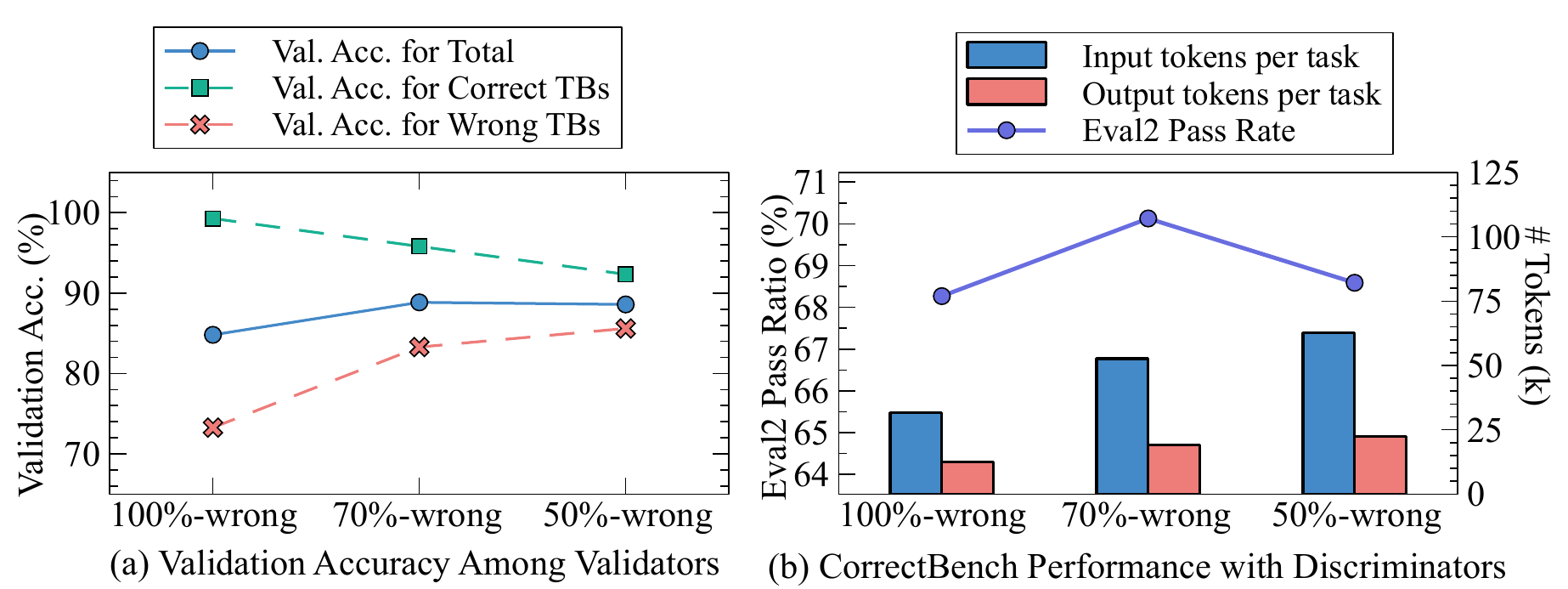}
    \vspace{-0.3cm}
    \caption{Comparison of validators}
    \label{fig: validators}
    \vspace{-0.7cm}
\end{figure}
As is mentioned in Section \ref{subsec 3-2: validator}, the validation criteria significantly influence the overall performance of \projectname{}. In this subsection, two sets of experiments are conducted to further explore the impact of different validation criteria from various perspectives, as depicted in \figname{} \ref{fig: validators}.

\figname{} \ref{fig: validators} (a) shows the validation (Val.) accuracy (Acc.) among different validators. To do this, we collected 1560 testbenches from the results of \cite{autobench} and ran the validators with different criteria (\textit{100\%-wrong}, \textit{70\%-wrong} and \textit{50\%-wrong}) on them. These testbenches are labeled with ``correct'' or ``wrong''. The definitions of the first two criteria are already elaborated in Section \ref{subsec 3-2: validator}, while the last criterion \textit{50\%-wrong} is similar to \textit{70\%-wrong} but only changed the percentage. These validators use the same RTL group, consisting of 20 correctness-unknown RTLs directly generated by \textit{gpt-4o-2024-08-06} for each task. If a validator generates the same result (\textit{``correct''} or \textit{``wrong''}) for a testbench as its label, then this validator is recognized as \textit{``success''} for this testbench. To better evaluate the performance of these validators, the validation accuracy for all testbenches, correct testbenches, and wrong testbenches are summarized, respectively, as shown in \figname{} \ref{fig: validators} (a). Evidently, with the validation threshold (the percentage) decreasing, the validation accuracy of recognizing correct testbenches also decreased. This is because the tendency to validate testbench as wrong is increasing. In other words, the validator is becoming more stringent in identifying erroneous testbenches. Consequently, the validation accuracy for identifying incorrect testbenches increased for the same reason.

Among the three criteria, \textit{70\%-wrong} achieves the highest global validation accuracy at 88.85\%, which is the criterion employed in our study. Although \textit{50\%-wrong} attains a comparable global validation accuracy, it has a lower validation accuracy for correct testbenches (92.34\%) compared to \textit{70\%-wrong}. A lower validation accuracy for correct testbenches implies a higher likelihood of specific tasks failing to converge. This could result in the validator never issuing a ``testbench pass" report for these tasks, thereby leading to further performance degradation of the entire system.

In addition to analyzing the existing data, we conducted a comparative experiment by implementing the entire framework using different validation criteria, as illustrated in \figname{} \ref{fig: validators} (b). The bars in the figure represent the token cost, while the points and line indicate the average performance across 156 Verilog tasks. The framework employing the \textit{70\%-wrong} validation criterion demonstrates the highest performance, which aligns with our previous analysis. Also, with the validator's intent to generate a ``testbench is wrong'' report, the total token cost is increasing because more such reports necessitate additional self-correction and rebooting iterations.

Due to the high cost of executing the entire workflow, only three criteria were compared in this work. Thus, the \textit{70\%-wrong} criterion utilized in our work may not be the optimal choice. Nonetheless, the limited experimental results already indicate the performance trends of the validators, with \textit{70\%-wrong} performing the best among the three criteria examined.

\subsection{Performance on Other LLMs}
\label{subsec 4-4: Performance on Other LLMs}
\begin{figure}
    \centering
    \includegraphics[scale=0.28]{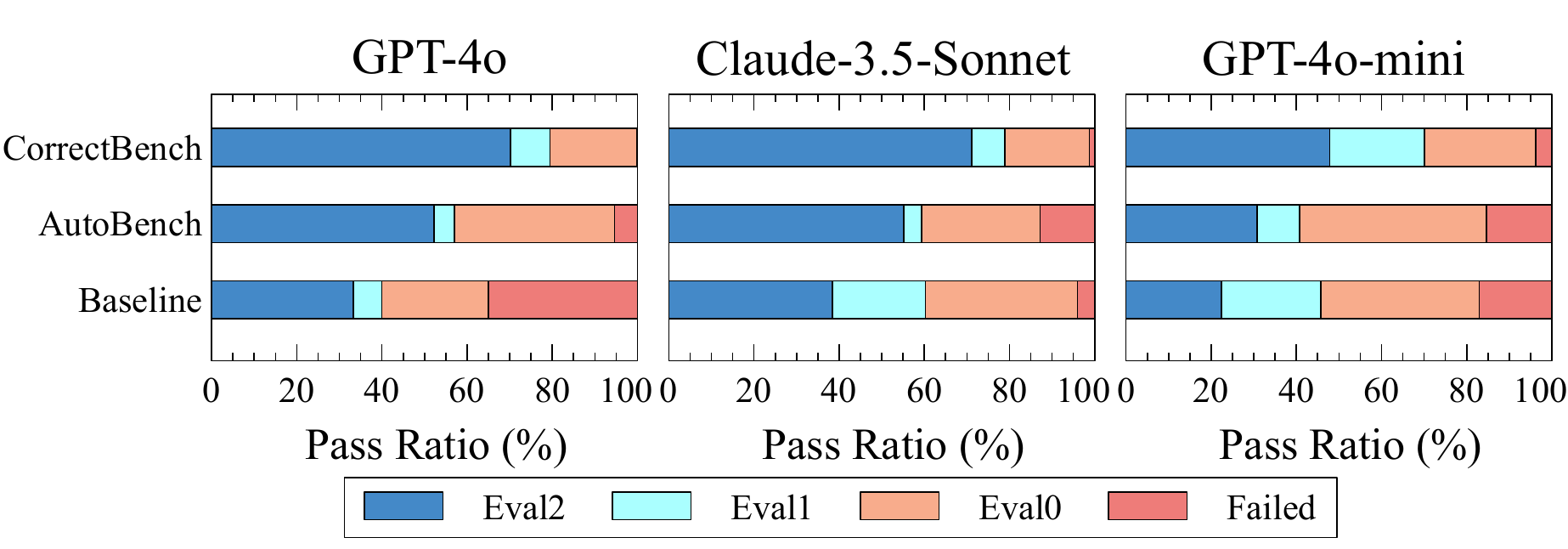}
    \vspace{-0.3cm}
    \caption{Performance of \projectname{} on Different LLMs.}
    \label{fig: LLMs comparison}
    \vspace{-0.5cm}
\end{figure}

To demonstrate that our workflow serves as a general methodology applicable to all the LLMs, we repeat the experiments outlined in Section \ref{subsec 4-2: Main Results} using two additional widely-used commercial LLMs: GPT-4o-mini (4o-mini) and Claude-3.5-Sonnet (Claude). Note that due to stricter daily token usage limitations, we conducted \projectname{} across 156 tasks on Claude only once. 
Furthermore, as the development of \projectname{} was conducted using \textit{GPT-4o}, its application on other LLMs might encounter format or interface compatibility issues, potentially leading to suboptimal results.

The comparison results are presented in \figname \ref{fig: LLMs comparison}. The blue bars illustrate the Eval2 pass ratios, where both \textit{Claude} and \textit{4o-mini} exhibit similar improvement among the methods. This indicates that our \projectname{} demonstrates consistent performance across these LLMs.

The performance of AutoBench in Eval1 and Eval0 on Claude and 4o-mini is occasionally inferior to the baseline. This can be attributed to the fact that Eval0 and Eval1 are not exhaustive metrics; the simpler testbenches generated by the baseline have a higher likelihood of avoiding syntax errors and reporting a ``pass" for the DUTs. However, these testbenches are not correct and consequently fail at Eval2.

\section{Conclusion}
\label{sec: conclusion}

In this work, we propose \projectname{}, the first automatic testbench generation framework with functional self-validation and self-correction. \projectname{} improved the generated testbench pass ratio to 70.13\%, compared with the previous work's 52.18\% and baseline's 33.33\%. Moreover, for sequential circuits, our work generates 66.18\% more correct testbenches than AutoBench and almost 5 times the baseline method.
Future research will explore the more advanced validation criteria, coverage-based self-validation, and extracting additional information to enable the corrector to perform more advanced correction.

\section*{Acknowledgment }

This work is funded by the Deutsche Forschungsgemeinschaft (DFG, German Research Foundation) – Project-ID 504518248 and by TUM International Graduate School of Science and Engineering (IGSSE).

\newpage
\let\oldbibliography\thebibliography
\normalem
\bibliographystyle{IEEEtran}
\bibliography{ref}

\end{document}